\documentstyle[11pt,epsfig]{article}

\newcommand{\kim}{ k_{1}^{\mu}}                                      
                                      
\newcommand{\ki}{ k_{1}}

\newcommand{\ko}{ k_{0}}

\newcommand{\p}{\partial}

\newcommand{\eps}{ \epsilon}

\newcommand{\kinb}{\mbox {$ \bar{k_{1}^{\nu}}$}}

\newcommand{\kib}{\mbox {$ \bar{k_{1}}$}}

\newcommand{\be}{\begin{equation}}
\newcommand{\br}{\begin{eqnarray}}
\newcommand{\ee}{\end{equation}} 
\newcommand{\er}{\end{eqnarray}}

\begin{document}
\renewcommand{\theequation}{\thesubsection.\arabic{equation}}

\title{
\hfill\parbox{4cm}{\normalsize IMSC/2004/04/18\\
                               hep-th/0404203}\\        
\vspace{2cm}
Holomorphic Factorization and Renormalization Group in Closed String Theory.
\author{B. Sathiapalan\\ {\em Institute of Mathematical Sciences}\\
{\em Taramani}\\{\em Chennai, India 600113}\\ bala@imsc.res.in}}           
\maketitle     

\begin{abstract} 
The prescription of Kawai, Lewellen and Tye for writing the
closed string tree amplitude as sums of products of open string 
tree amplitudes,
is applied to the world-sheet renormalization group equation.
The main point is that regularization of 
the Minkowski (rather than Euclidean)
world sheet theory
allows factorization into  left-moving and right-moving sectors
to be maintained. 
Explicit calculations are done for the tachyon and the 
(gauge fixed) graviton. 
\end{abstract}

\newpage
\section{introduction}

The renormalization group approach for  strings has been studied
for some time by many authors [\cite{L} - \cite{T}]. In particular for open strings,
because the calculation involves one-dimensional integrals,
a lot has been done. In \cite{BSPT} it was shown that a proper-time
equation for open strings can be written, which is essentially a 
Wilsonian renormalization group equation. It gives the full equation
of motion, unlike the $\beta$-function, which is only proportional to
the equation of motion. It was also shown that by keeping a finite cutoff
one can go off-shell and make contact with string field theory. This was further
made gauge invariant in \cite{BSLV} at the free level and a proposal
for the gauge invariant interacting theory was made in \cite{BSGI,BSREV}. 
It was also subsequently generalized to include Chan-Paton factors
in \cite{BSCP}. 
In order to generalize all this to closed strings we need a method
that allows us to use the open string loop variable techniques.

In \cite{KLT} Kawai, Lewellen and Tye (KLT) derived a prescription 
for writing down a closed string tree
amplitude as a the sum of  products of two open string amplitudes.
Closed string vertex operators are products of holomorphic and anti-holomorphic
vertex operators. The correlation functions therefore factorize into a product
of a holomorphic function and an anti-holomorphic function.
But  the S-matrix amplitudes involve  correlators integrated over the entire 
complex plane. The integration does not factorize (at least naively) and
the resultant closed string amplitudes do not appear to be directly related
to the open string ones. However, KLT showed that in fact, a Wick rotation
into Minkowski world-sheet can be performed and what is obtained is a product
of left and right moving correlation functions that are functions of
two real variables ($\sigma ~+~\tau$ and $\sigma - \tau$ respectively). 
The integrated amplitude also factorizes - except for a phase factor
that retains some correlation between the two sectors. The result is a sum
of terms, each of which is a product of open string amplitudes and a phase
factor.

The KLT technique is for on-shell S-matrix amplitudes. We would like to
get the corresponding equation of motion using the renormalization
group prescription. This involves introducing a regulator and then calculating the
$\beta$-function. The main challenge is to do this while maintaining
the  factorization property that KLT demonstrated for on-shell amplitudes
(which does not require a regulated world-sheet theory). This is the topic of this paper.

In order to derive a renormalization group equation the first step
is to regularize the theory so that there are no divergences. We do this
in Minkowski world-sheet rather that in Euclidean world-sheet.
The regulated propagator in Euclidean space is $ln~(z \bar z ~+~a^2)$.
This does not factorize into a holomorphic and antiholomorphic part.
 In Minkowski world sheet on the other hand, the propagator can be regulated
as $ln (\xi ^2 ~+~a^2) + ~ln (\eta ^2~+~a^2)$, which factorizes. This 
propagator is finite because $\xi , \eta$ are real after  Wick rotation. 
 There are still
infrared divergences because the variable $\xi , \eta$ are non-compact. We will
cutoff the integrals at $\pm R$ where $R \rightarrow \infty$. The infrared
divergences here have to be treated on the same footing as ultraviolet
divergences because poles in some channels are reflected in $ln ~a$ terms,
whereas poles in other channels show up as $ln ~R$ divergences. This can be
achieved by  setting $R ~=~{l^2 \over a}$. Effectively we have introduced a 
renormalization scale
$l$. The equations will involve the ratio $l\over a$.  
This renormalization scheme dependence is
 expected in off-shell
amplitudes. On shell they will disappear. 

 Another technique that has been used \cite{BSPT} is to restrict the integration region
by removing small portions around the singular point. This technique
cannot be used in the loop variable approach \cite{BSGI}. The loop variable
approach requires that one should be able to define more than one vertex operator
at a point.  This means that the propagator itself has to be regulated.
In this paper since we are more concerned with the KLT prescription as
a way of possibly implementing the loop variable approach, we 
use the regulated propagator. But for simplicity, as will be clear later,
 we will also remove
small portions of the contour of integration.

We will use the same techniques as KLT to derive the phase factor in the
cutoff theory, i.e. we start with the Euclidean world sheet and analytically
continue to Minkowski world-sheet, taking care not to cross any singularities.
 We assume the same propagator in the Euclidean version.
The propagator is thus $ln ~ (z^2 ~+~ a^2)(\bar z ^2 ~+~ a^2)$. While this
amounts to a modification of the short distance structure, the theory is
not regulated. Nevertheless in the limit $a\rightarrow 0$ we expect to
recover the correct S-matrix for on-shell states. We then analytically continue
to Minkowski space, where this theory is finite. Again as $a\rightarrow 0$ we
expect to recover the S-matrix provided during the Wick rotation we do not
cross any singularities. As in the $a=0$ case of KLT one gets a prescription
for the contours and phase factors. In fact the result for the phase factors
is exactly the same - it does not depend on $a$.

Once we have introduced a regulator, in principle the external 
momenta can be taken off-shell (i.e. need not satisfy the physical state conditions).
 For on-shell (physical) states the cutoff can be taken to zero
and we recover the usual amplitudes.
We can use this to obtain an R-G equation by studying the $a$-dependence.
If the equation of motion is satisfied we expect $d/d~ln~a ~=~0$. If we want
the coefficient of the leading log, we can set $l=a$ at the end. 
The crucial question for off-shell amplitudes is whether it can be made
gauge invariant. This will not be discussed here. It is likely that
the loop variable techniques used for the open string can be applied here
also.  This will be discussed elsewhere \cite{BSC}.

This paper is organized as follows. In section 2 we describe the
 KLT prescription
as applied to the theory with a cutoff. In section 3 we apply it to get the
quadratic and cubic terms in the tachyon equation (cubic and quartic
 terms in the
Lagrangian) and also the cubic three-graviton vertex. 
Section 4 contains some concluding remarks.

\section{KLT Prescription}

Let us consider the integral 
\be   \label{T1}
\int ~d^2z ~ |z|^{2\alpha}
\ee
  We write
$z~=~x~+~iy$ and analytically continue in $y ~= ~y'~+~iy''$ : Instead
of integrating along the real $y$-axis ($y'$) we continue to the imaginary
axis ($y''$). Thus $z$ becomes $x~-~y'' \equiv ~\xi$ . The old and new 
contours are depicted in Figure 1.  Similarly $\bar z$ becomes $x~+~y''
\equiv ~\eta$. These are of course nothing but the Minkowski space
light cone coordinates - left moving and right moving respectively.
The philosophy is that the Euclidean contour which is known to give the correct
amplitude is continued analytically to Minkowski space, taking care not
to cross any singularities. The singularities (branch points)
 are shown in Figure 1. 
They correspond to $\xi ~ =~0$ ('A') and $\eta ~=~0$ ('B'). At A, $y'' ~=~ x$
and therefore  (at A) $\eta ~=~ 2x$. Thus for $\eta > 0$ A is on the upper
half plane. Similarly the point B is $y'' ~=~-x$.  The rotated contour 
C' is shown in the figure. If we draw this contour in the $\xi$ plane
we have the situation shown in Figure 2. Note that as $y''$ increases,
$\xi$ decreases. When $\eta >0$ which is the case in Figure 1,
as you go along the contour the branch point is on your left. This is
shown in Figure 2. If $\eta <0$ we have the situation in Figure 3,
the branch point is to the right of the contour. 
One can similarly draw contours
for $\eta$ by studying the branch point B. This is Figure 4,5. 
\begin{figure}[htbp]
\begin{center}
\epsfig{file=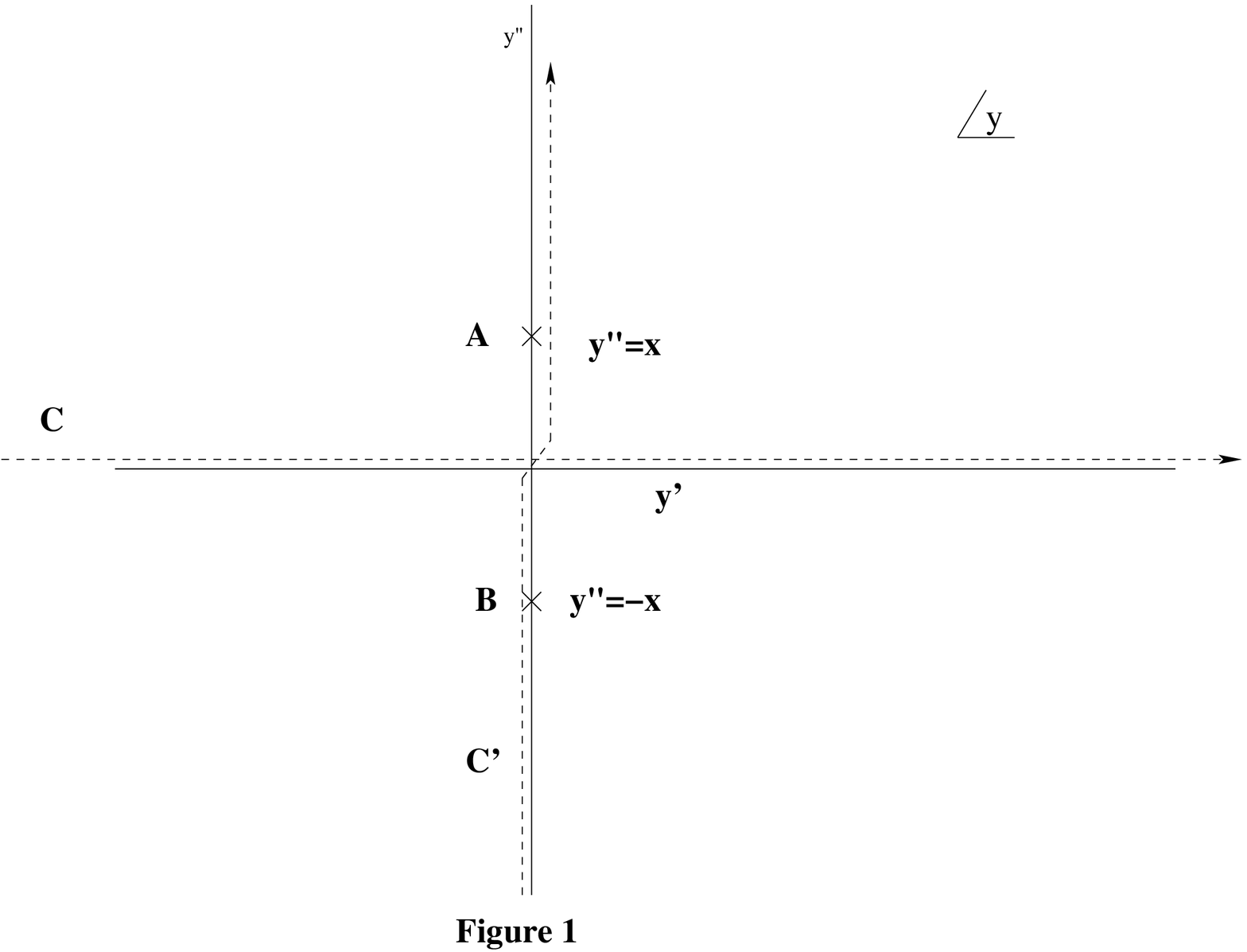, width= 8 cm,angle=0}
\vspace{ .2 in }
\begin{caption}
  {Wick rotation of contour from C to C'. The  singularities at
A and B are to be avoided. }
\end{caption}
\end{center}
\label{Figure 1}
\end{figure}

\begin{figure}[htbp]
\begin{center}
\epsfig{file=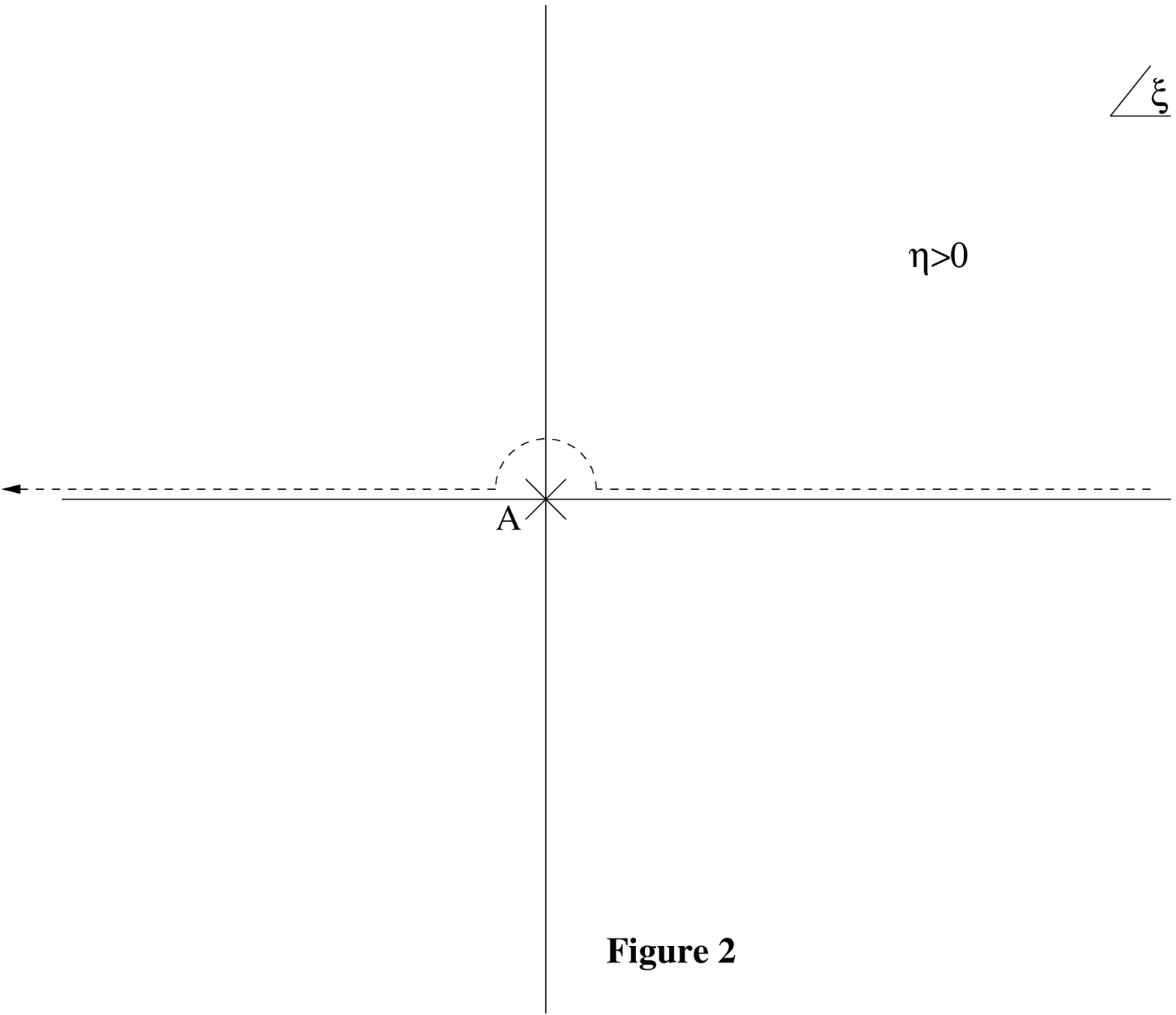, width= 7cm,angle=0}
\vspace{ .2 in }
\begin{caption}
  {Contour in $\xi$-plane starts at $+\infty$ and goes left.
 The singularity A is to the left of contour just as in Figure 1.
 This is for $\eta >0$. }
\end{caption}
\end{center}
\label{Figure 2}
\end{figure}

\begin{figure}[htbp]
\begin{center}
\epsfig{file=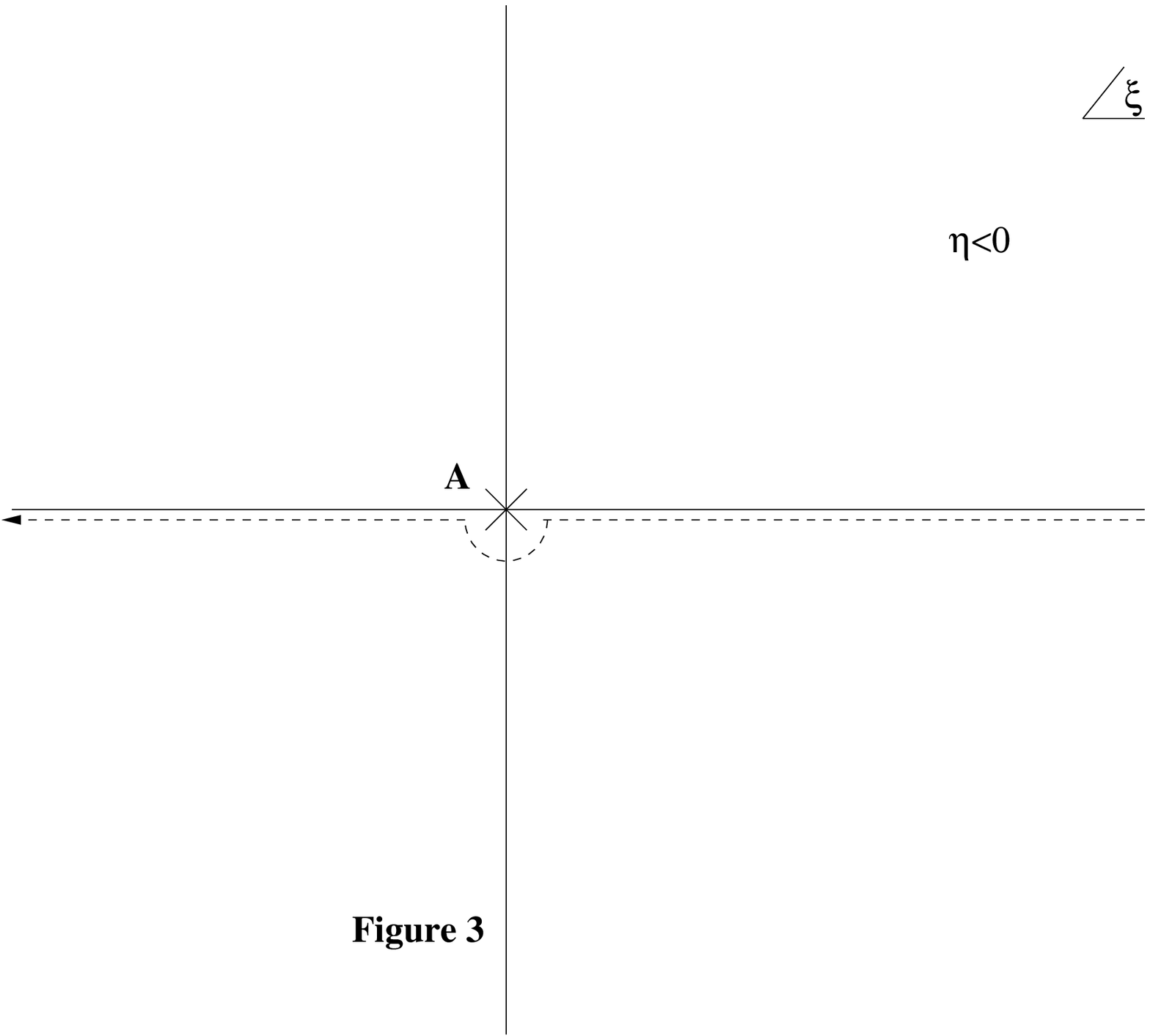, width= 7cm,angle=0}
\vspace{ .2 in }
\begin{caption}
  {Contour in $\xi$-plane for $\eta <0$. }
\end{caption}
\end{center}
\label{Figure 3}
\end{figure}
\begin{figure}[htbp]
\begin{center}
\epsfig{file=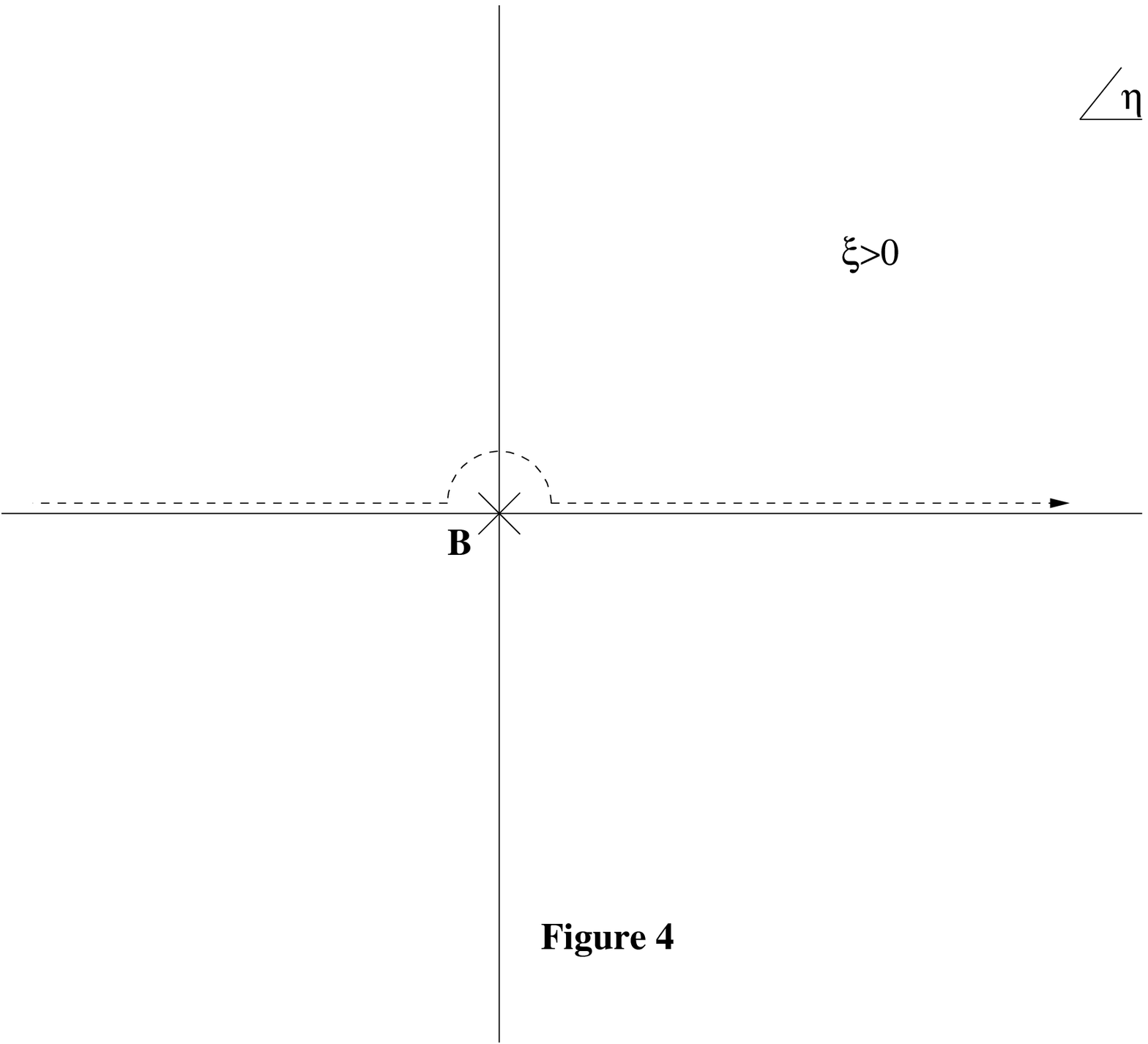, width= 7cm,angle=0}
\vspace{ .2 in }
\begin{caption}
  {Contour in $\eta$-plane starts at  $-\infty$ and goes from left to right. 
For $\xi >0$ the singularity B is to the right of the contour. }
\end{caption}
\end{center}
\label{Figure 4}
\end{figure}

\begin{figure}[htbp]
\begin{center}
\epsfig{file=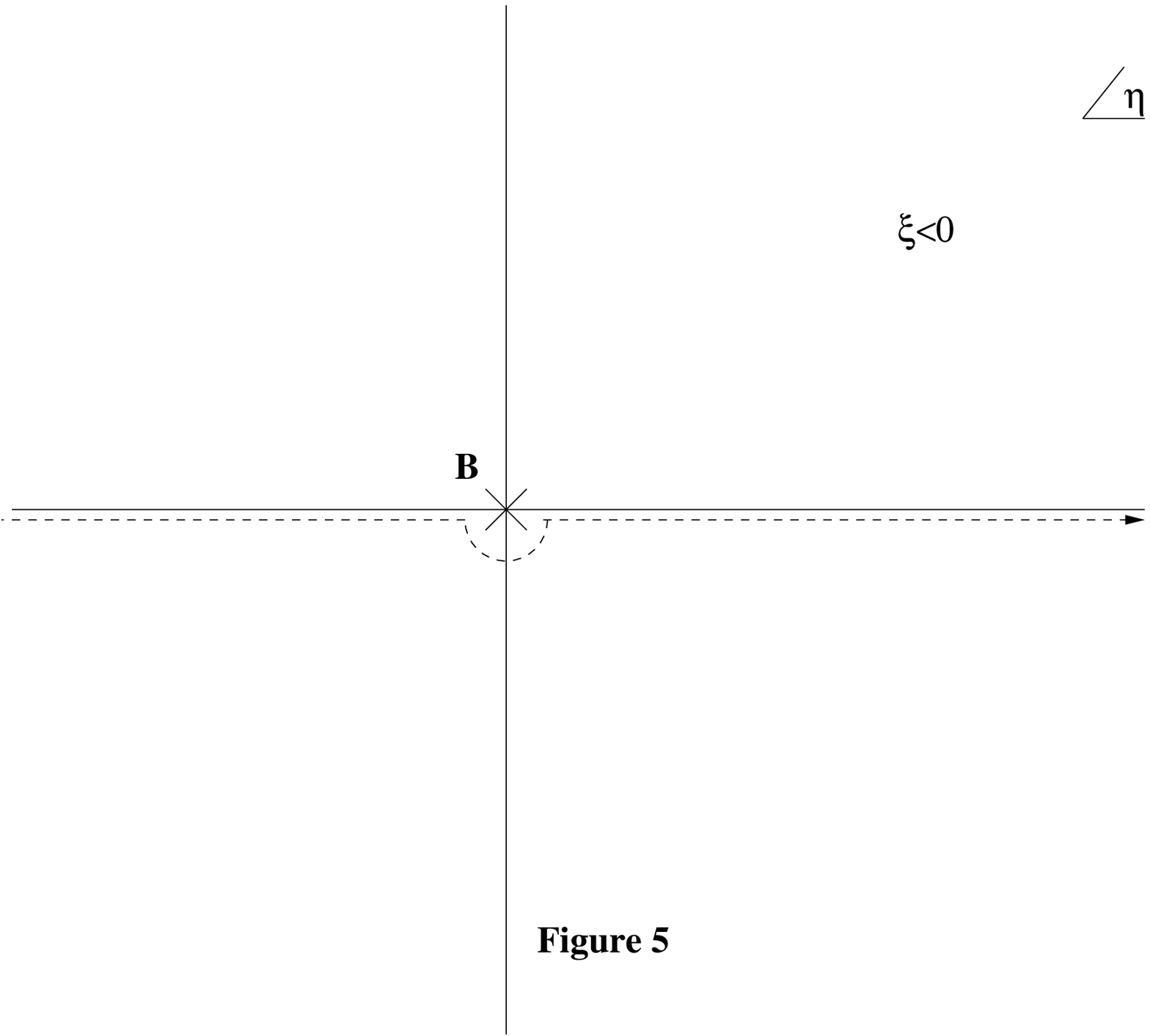, width= 7cm,angle=0}
\vspace{ .2 in }
\begin{caption}
  {Contour in $\eta$-plane starts at $-\infty$ and goes to the right.
 The singularity B is to the left of contour for $\xi <0$ }
\end{caption}
\end{center}
\label{Figure 5}
\end{figure} 
From this one can deduce the phases:
\br \label{Ph}
1. ~~~~~~~~~
 \xi  > 0,  \eta  >0 & \Rightarrow & \xi  =  |\xi | ,~~~ \eta = |\eta |
\nonumber \\
2. ~~~~~~~~~
 \xi >0, \eta <0 & \Rightarrow &  \xi = |\xi | , ~~~ \eta = |\eta | e^{i\pi}
\nonumber \\
3.  ~~~~~~~~~
\xi < 0 , \eta >0 & \Rightarrow & \xi = |\xi | e^{i\pi} , \eta = |\eta |
\nonumber \\
4. ~~~~~~~~~
\xi <0 ,\eta <0 & \Rightarrow & \xi = |\xi | e^{i\pi} , \eta = |\eta | e^{-i\pi}
\er

Case 4 is obtained from 3 by continuing $\eta$. We can also get a different
 phase by starting
from 2 and analytically continuing $\xi$. However the total phase
 of the product $\xi \eta$ is unaltered.
It is $\pi$ in case 2,3 and $0$ in case 1,4. Thus the phase factor of $\xi \eta$ 
 can be expressed as 
$e^{i\pi \theta (-\xi \eta )}$. $\theta $ is the usual step function. The integral
in (\ref{T1}) thus becomes 
\be    \label{T1.1}
\int _{-\infty}^{\infty} ~d\xi ~\int _{-\infty}^{\infty}
 ~d\eta ~|\xi |^\alpha |\eta |^\alpha ~e^{i\alpha \pi \theta (-\xi \eta )}
\ee 

When there are several variables this has a simple generalization:
 If $\xi _i, \xi _j,
\eta _i, \eta _j$ are the variables, the phase factor becomes
 $e^{i\pi \theta (-(\xi _i - \xi _j)(\eta _i -\eta _j))}$. 
For each pair there is such a factor. This is the KLT prescription.

If the integral is of the form 
\be   \label{T1.2}
\int _{-\infty}^{\infty} ~d\xi ~\int _{-\infty}^{\infty}
 ~d\eta ~\xi ^{\alpha _1} \eta ^{\alpha _2} 
\ee 
with $\alpha _1 \neq \alpha _2$, then there is an ambiguity in the phase
depending on whether we reach case 4 from case 3 or from case 2.
So in this case we need a prescription. Let us assume that we follow the 
prescription 1-3-4, i.e. $\xi$ is always continued before $\eta$ in order to
reach case 4. Then we get a phase 
$|\xi | ^\alpha _1 |\eta | ^\alpha _2 e^{i\pi(\alpha _1 -\alpha _2)}$ for
case 4. This prescription dependence will show up when we regulate the theory
in order to go off-shell (see eqn (\ref{T7})).

A simple regularization prescription is to cutoff the integration
region around the origin. Thus we write (\ref{T1.1}) as 
\be    \label{T2}
[\int _{a} ^{\infty} d\xi ~+~ \int  _{-\infty} ^{-a} d\xi ]~ 
[\int _{a} ^{\infty} d\eta ~+~ \int  _{-\infty} ^{-a} d\eta ]~ 
|\xi |^\alpha |\eta |^\alpha ~e^{i\alpha \pi \theta (-\xi \eta )}
\ee

If we assume that $1+\alpha < 0$, we do not need an infrared regulator.
We get $2{a^{2\alpha +2}\over (\alpha +1)^2}[1-e^{i\pi (\alpha +1)}]$.
In the limit $1+\alpha \rightarrow 0$ this becomes 
$2i\pi {a^{2\epsilon}\over \epsilon}$,
where $\epsilon = 1+\alpha$. 

Another regularization prescription, more suited for the loop variable approach
of \cite{BSGI} is to replace the propagator in Minkowski space by 
$ln~ (\xi ^2 + a^2) + ln ~ (\eta ^2 + a^2)$. In this case the contours 
in the $y$-plane and $\xi$-plane  are as shown
in Figure 6,7. A convenient prescription is to integrate from $+\infty $ to $0$ 
('a' to 'b') and
$0$ to $-\infty$ ('d' to 'e'), i.e. we drop the portion 'bcd' in Figure 7.
 In the limit $a\rightarrow 0$
this contribution is zero anyway. 
If we are concerned with off-shell amplitudes, so that
$a$ is finite, the this will modify the answers somewhat.
 On-shell both are equivalent.
The method of regularizing the propagator has the advantage that it can easily
 be made gauge invariant
using the loop variable approach.
In the absence of further criteria to pick one off-shell prescription
 over another we use this
one. 
\begin{figure}[htbp]
\begin{center}
\epsfig{file=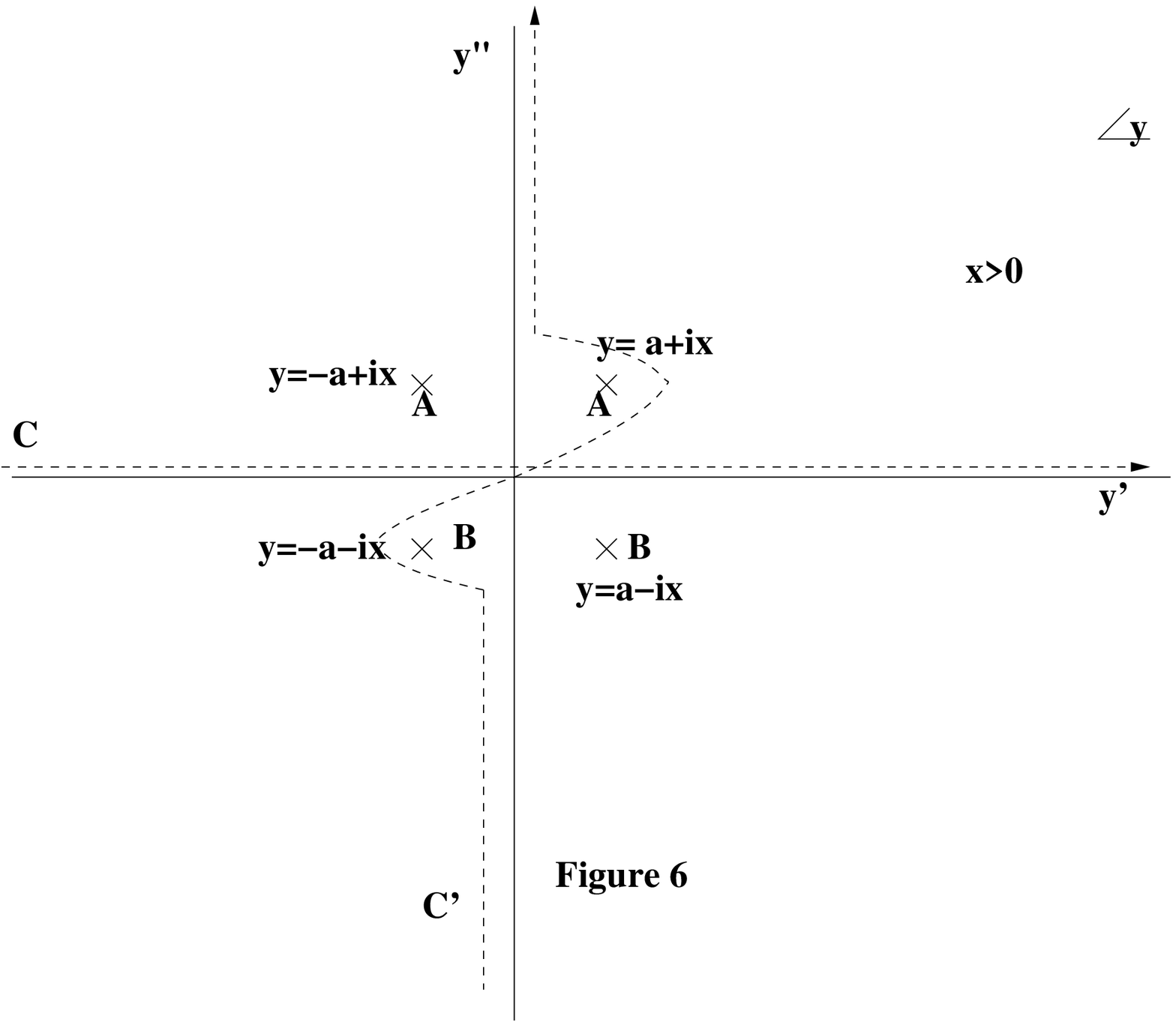, width= 7cm,angle=0}
\vspace{ .2 in }
\begin{caption}
  {Contour C in $y$-plane is rotated to C' avoiding  singularities A and B.
 The singularity A is to the left of contour for $x>0$. }
\end{caption}
\end{center}
\label{Figure 6}
\end{figure}
\begin{figure}[htbp]
\begin{center}
\epsfig{file=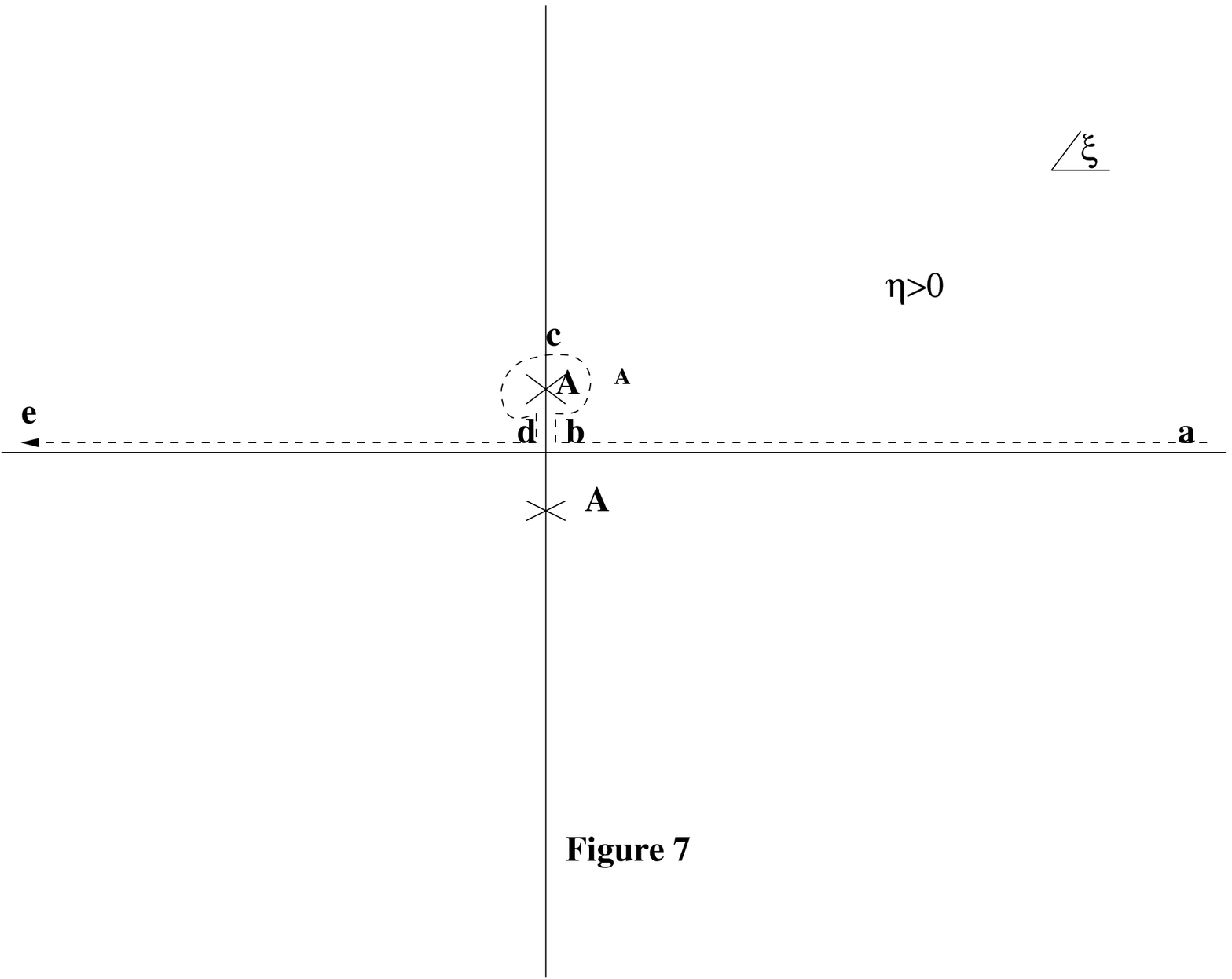, width= 7cm,angle=0}
\vspace{ .2 in }
\begin{caption}
  {Contour C in $\xi$-plane. 
 The singularity A is to the left of contour for $\eta >0$. The regularization prescription is 
to drop the contribution of section 'bcd' of the contour. }
\end{caption}
\end{center}
\label{Figure 7}
\end{figure}

In this prescription (\ref{T2}) is replaced by
\be \label{T3}
[\int _{0} ^{\infty} d\xi ~+~ \int  _{-\infty} ^{0} d\xi ]~ 
[\int _{0} ^{\infty} d\eta ~+~ \int  _{-\infty} ^{0} d\eta ]~ 
(\xi ^2 + a^2 )^{\alpha \over 2} (\eta ^2 + a^2) ^{\alpha \over 2}
 ~e^{i\alpha \pi \theta (-\xi \eta )}
\ee

The integral $\int _0^\infty ~d\xi ~ (\xi ^2 + a^2)^{\alpha \over 2} ~=~ 
{1\over 2}(a)^\epsilon ~
 {{\Gamma (-{\epsilon \over 2})\Gamma ({1\over 2})}\over
 {\Gamma ({{1-\epsilon } \over 2})}}$.
Here, as before, $\epsilon = 1+\alpha$. When $\epsilon \rightarrow 0$
 we get for the integral
$-{a^\eps \over \eps}$. The final expression for (\ref{T3}) in this limit 
is exactly the same as for (\ref{T2}).  We will now apply all this to the tachyon
and graviton. 

\section{$\beta$-function}

\subsection{Tachyon}

The tachyon vertex operator is $\int ~d^2z~e^{ik.X}$ and in conventions where the
propagator $<X ^\mu (z,\bar z) X ^\nu (w, \bar w) > ~=~ -{g^{\mu \nu}\over 2}
 ln~|z-w|$
($g^{00} = -1$), 
the dimension of this vertex operator is $k^2\over 4$. We want the dimension to be 2
and this gives the mass shell condition $k^2=8$.  Equivalently if we normal order
we get $e^{ik.X} = :e^{ik.X}: a^{k^2\over 4}$. The vertex operator
is now $\int ~{d^2z\over a^2} :e^{ik.X}: a^{k^2\over 4}$. We have introduced
explicit powers of $a$ to compensate for  $d^2z$. If we require 
${d\over d~ln~a} =0 $ we get the equation $k^2 = 8$.     

At the next order we have
 ${1\over 2!}\int ~ d^2 z_1 \int d^2 z_2 <:e^{ik_1.X(z_1) }:~:e^{ik_2. X(z_2)}:>$
This boils down to the integral $\int d^2 z |z|^{k_1.k_2 \over 2}$, where we have
used $z = z_1-z_2$. This corresponds to (\ref{T1}) with $\alpha = {k_1.k_2 \over 4}$.
When the two incoming tachyons and the ``outgoing'' tachyon with momentum $k_1+k_2$,
are on shell, i.e. $k_1^2=k_2^2= (k_1 +k_2)^2 =8$ 
we have the condition $1+\alpha \equiv \eps =0$ and using the result of (\ref{T2}) we get
$2\pi i a^{2\eps}\over \eps$.  Thus when we do $d\over d~ln~a$ we get $4i\pi$
as the coefficient of the quadratic term in the equation of motion (the precise
normalization will not concern us here).

Now consider the cubic term in the equation of motion (quartic term in the action).
We will fix one vertex operator at $z_1~=~0$. Thus $\xi _1 ~=~ \eta _1 ~=~0$.
We thus have to evaluate
\[
\int _{-R}^{+R}~ d\xi _3 ~\int _{-R}^{+R}  ~d\xi _2 
(\xi _3 -\xi _2)^{k_2.k_3} (\xi _3)^{k_3.k_1}(\xi _2)^{k_2.k_1} 
\]
\[
\int _{-R}^{+R}~ d\eta _3 ~\int _{-R}^{+R}  ~d\eta _2 
(\eta _3 -\eta _2)^{k_2.k_3} (\eta _3)^{k_3.k_1}(\eta _2)^{k_2.k_1}  
\]
\be   \label{T4}
e^{i\pi k_1.k_2 \theta (-(\xi_1-\xi _2)(\eta _1-\eta _2))~+~
i\pi k_1.k_3 \theta (-(\xi_1-\xi _3)(\eta _1-\eta _3))~+~
i\pi k_3.k_2 \theta (-(\xi_3-\xi _2)(\eta _3-\eta _2))}
\ee

To get the off-shell answer we have to regulate divergences. 
The divergences at $\pm \infty$
are regulated by $R$. The other divergences can be regulated
 either by cutting 
off the integration region or using the regulated propagator.
 If we are concerned
about gauge invariance we should use the regulated propagator. In this paper
we will use the simpler prescription of cutting
 off the region of integration. In an appendix we compare the two regularization 
schemes for a typical integral of this type.
 \footnote{The leading terms relevant for the on-shell
calculation are seen to be the same (as expected) in both schemes.}
Also
we will simplify our calculation by keeping our external particles
 close to the mass shell
so that we can use all the simplifications of the on-shell calculation. 
This means that
all terms in the equation of motion that vanish when the external
 fields are on-shell,
 are dropped.  This allows us to get away with evaluating
 fewer contour integrals. However if we want to go off-shell using the proper-time
equation, one cannot do this.

In evaluating the integrals in (\ref{T4}) the simplest procedure is to
 fix the ordering
of the $\xi$'s and $\eta$'s and use the fact that whenever
 $\eta _i ~>~\eta _j$, the 
$\xi _i$ contour goes above the branch point at $\xi _i = \xi _j$. 
Now we have two possibilities: I)  $\eta _3 ~> ~0$  and II) $\eta _3 ~<~0$.

In each case there are {\it a priori} four ordering possibilities: 
\begin{enumerate}
\item  $\eta _2 ~>~ \eta _3$ and $\eta _2 ~>~0$ 
\item  $\eta _2 ~<~ \eta _3$ and $\eta _2~>~0$ 
\item $\eta _2 ~>~ \eta _3$ and $\eta _2 ~<~0$ 
\item  $\eta _2 ~<~ \eta _3$ and $\eta _2~<~0$
\end{enumerate}

Of course in case I ordering 2 is not possible and in case II ordering 3
 is not possible.

For each ordering the contours are shown in Figure 8.
 (In the figure we have $\xi _3 >0$.)
\begin{figure}[htbp]
\begin{center}
\epsfig{file=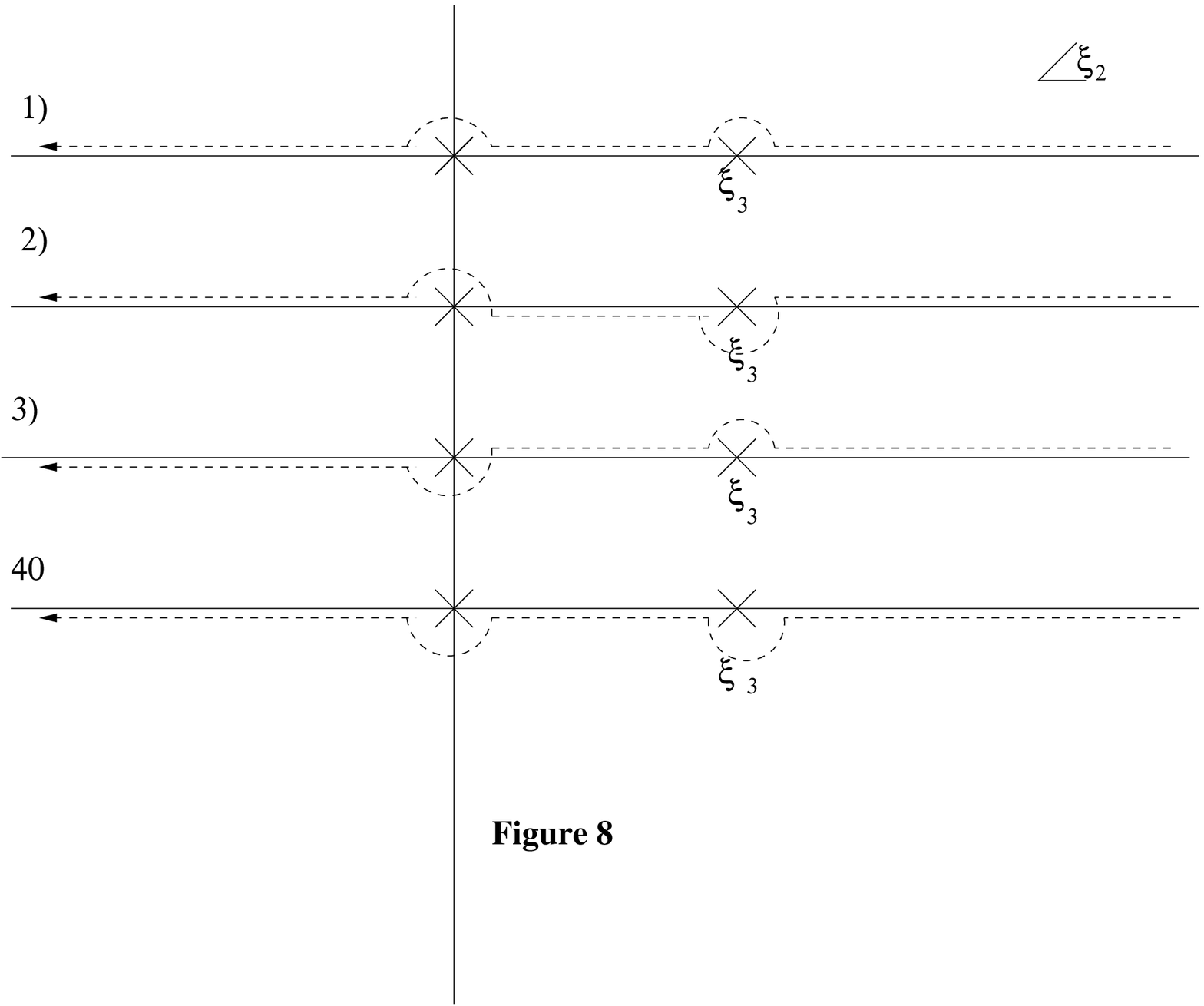, width= 10cm,angle=0}
\vspace{ .2 in }
\begin{caption}
  {Contours  in $\xi _2$-plane corresponding to the cases 1-4 listed above. 
The singularities are
at $\xi _2 = \xi _1 =0$ and $\xi _2 = \xi _3$}
\end{caption}
\end{center}
\label{Figure 8}
\end{figure}

It is clear that contours 1 and 4 will not contribute since they
 can be closed without enclosing
any singularities. Thus we have contour 3 for case I and contour 
2 for case II. 
They can both be deformed
to the contour shown in Figure 9.

\begin{figure}[htbp]
\begin{center}
\epsfig{file=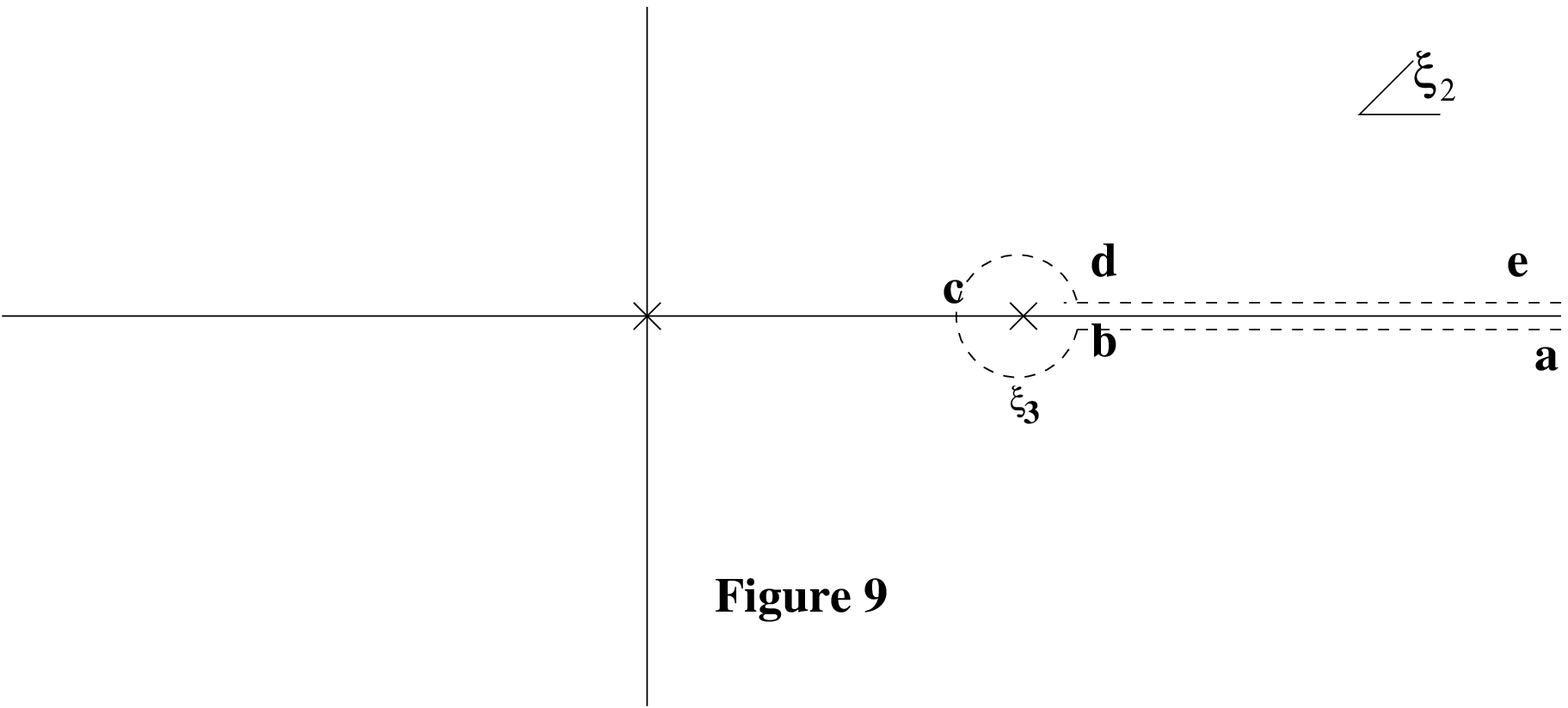, width= 10cm,angle=0}
\vspace{ .2 in }
\begin{caption}
  {Contours in case 2 and 3 of figure 8 can be deformed to this contour.}
\end{caption}
\end{center}
\label{Figure 9}
\end{figure}

 Our strategy will be to 
regulate the final contour integral along the contour shown in Figure 9. 
This will be done by removing a small circle around the branch point.
We are guaranteed that it reduces to the S-matrix calculation when the regularization
is removed. The only remaining issue is whether this is gauge invariant. This requires
the loop variable techniques of \cite{BSGI}. Note that if instead of regulating the theory 
after closing the contours at $\infty$ and dropping the ones that do not enclose singularities,
we were to regulate all integrals from the beginning,
the answers would be different. This is because integrals that are zero
in the S-matrix calculation and therefore have been dropped entirely
are no longer zero - simply because small semi-circles have been removed
from them and that modifies the answer from zero to non-zero. However one expects that
in the on-shell limit which is the continuum limit on the world sheet, these contributions
will drop out. Thus the off-shell equations depend on the prescription employed.  
 It is also  conceivable that one has
to include all the four contours in Figure 8 when one uses the loop variable approach.
This remains to be investigated.

In Figure 9 we have to evaluate $\int _a^b + \int _d ^e$. Both integrals
are equal up to a phase to
\be
(\xi _3) ^{k_{13}+k_{12}+k_{23} +1} [B(1+k_{32},1+k_{13}) 
-{({\xi _3 \over R})^{1+k_{13}}\over (1+k_{13})}
- {({a\over \xi _3})^{1+k_{23}}\over (1+k_{23})}]
\ee 
The phases are $e^{-i\pi k_{23}}$ and $-e^{i\pi k_{23}}$. 
 They add to give a factor
$-2i~ sin~\pi k_{23}$. Thus the net result is

\be   \label{T5}
(\xi _3) ^{k_{13}+k_{12}+k_{23} +1}(-2i~ sin~\pi k_{23}) [B(1+k_{32},1+k_{13}) 
-{({\xi _3 \over R})^{1+k_{13}}\over (1+k_{13})}
- {({a\over \xi _3})^{1+k_{23}}\over (1+k_{23})}]
\ee
The regulated $\eta _2$ integral is
\be
\int _a^{\eta _3 -a} ~d\eta _2 (\eta _3 - \eta _2)^{k_{23}}
(\eta _2) ^{k_{21}} (\eta _3)^{k_{31}}
\ee 
This is 
\be      \label{T6}
(\eta _3)^{k_{13}+k_{12}+k_{23} +1}[B(1+k_{32},1+k_{12})
 -{({a\over \eta _3})^{1+k_{12}}\over (1+k_{12})}
- {({a\over \eta _3})^{1+k_{23}}\over (1+k_{23})}]
\ee

The rule for $\xi _3$ contour is the same: when $\eta _3 ~<~0$,
 choose the contour that goes above the singularity
at $\xi _3 ~=~0$, and vice versa. The unregulated $\xi _3$ integral would 
just be zero since one can close the contour at infinity.
  However, regularization
 removes the small semi-circle (of radius $a$) around the origin.
  Thus the value of the required integral
is simply the negative of the value of the integral around the small semi-circle.
 Thus the integrals are of the form
 $\int _{-a}^{+a} ~d\xi _3 \int _a^R~d\eta _3$,
where the integrands are given in (\ref{T5},\ref{T6}).

There are many terms to be integrated but they are all of the type:
\be \label{T7}
\int _{-a}^a ~d \xi _3 \int _a ^R ~d \eta _3 
~\xi _3 ^{-1+\eps} \eta _3 ^{-1+\eps '} R^x a^{-\eps -\eps ' -x} \times c
\ee

The pole term is signalled by $ln ~a$.  Thus we keep only those terms
 that can contribute $a^\delta$ where $\delta \approx 0$.
If $\delta$ is finite and greater than zero, then this term goes to zero
 as $a\rightarrow 0$. By analyticity
of the physical amplitude in momenta, we take this term to be zero for all
 non-zero $\delta$. At $\delta ~=~0$
there is a pole and we are attempting to remove that. It is in this sense
 that we must understand the limit $a\rightarrow 0$. Furthermore by dimensional
analysis, the final answer must be a function of $R/a$. Thus the coefficient
of $ln~a$ is always the same as the coefficient of $ln ~R$. If we set
$R={l^2\over a}$ an additional factor of $2$ is obtained for the coefficient of $ln~a$. 

Note also that (\ref{T7}) is not of the form given in (\ref{T1.1}) because
$\eps$ and $\eps '$ are in general not equal. Thus there is an ambiguity
in the phase prescription which is $e^{ \pm i(\eps - \eps ')\pi \theta ( -\xi \eta )}$ 
(see discussion after (\ref{Ph})). Of course in the limit that 
$\eps ,\eps ' \rightarrow 0$ this ambiguity disappears. This is the case
in the usual $\beta$- function calculation where on-shell intermediate
states are being removed. This is what is being attempted in this paper.
However if we are far off-shell, and with a finite
cutoff this ambiguity has to be dealt with by adopting a prescription. 
One such prescription was given in the paragraph after (\ref{T1.2}).
Perhaps fortuitously,
in the loop variable approach, because
the propagators are Taylor expanded in (integral) powers of $\eta , \xi$
we never see fractional powers of the kind we have in (\ref{T5}),(\ref{T6}).
With integer powers there is no ambiguity. So we do not have this problem
in the loop variable approach.  
 
In (\ref{T7}) the $\xi _3$ integral gives $i\pi a^\epsilon $
 (for $\eps \approx 0$). The $\eta _3$ integral
gives $(R^{\eps '} -a^{\eps '})\over {\eps '}$. If we extract the 
coefficient of $ln ~a$ in the product
$i\pi c a^\eps (R^{\eps '}-a^{\eps '})a^{-\eps -\eps ' -x}R^x \over \eps '$
 we get $-i\pi c$. The point to note
is that it does not depend on any of the parameters $\eps,\eps ',x$. 
Thus if we perform the operation
$d\over d~ln ~a$ and evaluate at $R=a$ (this extracts the $ln ~a$ part) 
 we get the S-matrix with it's pole parts removed, as the coefficient
 of the quartic term in the tachyon action:
\be
sin (\pi k_{23})B(1+k_{23},1+k_{13})B(1+k_{23},1+k_{12}) ~-~ pole~~parts
\ee

All this is of course exactly as in the case of the open string.

\subsection{Graviton}

Apart from the complication of extra indices the three graviton vertex
calculation is the same as the tachyon calculation. 
\footnote{We are doing the gauge fixed case here. The gauge invariant case
will be done using loop variables \cite{BSC}}.
So we will be very brief. The graviton vertex operator can be written as
\be
\int ~d^2z~ \kim \kinb :\p _z X^\mu \p _{\bar z} X^\nu e^{i\ko .X}:
\ee

$ \kim \kinb \approx h^{\mu \nu}$ represents a transverse traceless massless graviton.
Thus the physical state conditions are $\ko ^2 = 0 = \ki .\ko = \kib .\ko = \ki .\kib$. 

The quadratic term in the equation of motion is obtained by evaluating
\be
\int ~d^2z~ \kim \kinb :\p _z X^\mu \p _{\bar z} X^\nu e^{i\ko .X}:
\int ~d^2w~ p_1^\rho \bar p _1^\sigma :\p _w X^\rho \p _{\bar w} X^\sigma e^{ip_0 .X}:
\ee
as an OPE and extracting the part proportional to a graviton vertex operator.

We will only look at one of the many contractions possible.  Keeping in mind that
\br
< \p _z X^\mu (z) X^\nu (w) > &=& -{g^{\mu \nu}\over 4 (z-w)}\\ \nonumber
<  X^\mu (z) \p _w X^\nu (w) > &=& {g^{\mu \nu}\over 4 (z-w)}
\er
one of the terms is:
\be   \label{3G}
[\kim p_1^\rho {1\over 16|z-w|^2}p_0^\mu \kinb \bar p_1^\sigma \ko ^\sigma]
 :\p _w X^\rho \p _{\bar z} X^\nu e^{i\ko X(z) + p_0 X(w)}:|z-w|^{\ko .p_0\over 2}
\ee
If we Taylor expand $X(z)$ about $X(w)$, we get a graviton vertex operator 
$\p _w X ^\rho \p _{\bar w} X^\nu e^{i(\ko + p_0).X(w)}:$. 
The integral is of the form (\ref{T1}) with $\alpha = {{\ko .p_0 \over 4} -1}$.
Thus the result of doing the integral is $2\pi i a^{2\eps}\over \eps$ and $d\over d~ln~a$ 
acting on it gives $4\pi i$ as the coefficient of the leading log. As for the index
structure, if we think of this
as a vertex operator for a graviton field $h^{\rho \nu} \approx q_1^\rho \bar q_1^\nu $
then the three graviton coupling implied by the equation of motion (\ref{3G}) is:
\be
(q_1^\alpha \kim p_1^\rho)~ (\bar q_1^\beta \kinb \bar p_1 ^\sigma) 
~[\eta ^{\alpha \rho } p_0^\mu] [\eta ^{\beta \nu} \ko ^\sigma] 4\pi ~i
\ee
One can check that this agrees
 (up to overall normalization, which we are not concerned with at the moment)
with results available in the literature
\cite{GSW}.  The other terms in the three graviton coupling are related
 to this by symmetry.  This concludes our discussion of the $\beta$-function 
calculation for the tachyon and graviton.

\section{Conclusions}

We have shown how the KLT prescription can be applied
to the RG equation on the closed string world sheet. The main
point of the construction is that we regularize the Minkowski world
sheet theory rather than the Euclidean one. This allows us to maintain
the factorization of the amplitude into holomorphic and anti-holomorphic 
parts. 

The main motivation for attempting this left-right factorization is that
this gives the possibility of applying the loop variable techniques of \cite{BSGI}
to make the closed string RG equation gauge invariant. 
We hope to report on this soon.

{\bf Acknowledgements:} I would like to thank G. Date for useful discussions.

\appendix
\section{Appendix}

In this appendix we compare the two regularizations of the integral:
\be
\int _0 ^1 ~dx (1-x)^{q-1}x^{p-1} ~=~ B(p,q)
\ee

The first one is 
\[
\int _a ^{1-a} ~dx (1-x)^{q-1}x^{p-1} ~=~  
\int _0 ^1 ~dx (1-x)^{q-1}x^{p-1} - \int _0 ^a ~dx (1-x)^{q-1}x^{p-1} -\int _{1-a} ^1 ~dx (1-x)^{q-1}x^{p-1} 
\]
\be
=~B(p,q) - B_a(p,q) - B_a (q,p)
\ee
$B_a(p,q)$ is the incomplete Beta function with the expansion \cite{GR}
\be
B_a(p,q) = {a^p\over p}[1+{p(1-q)\over (p+1)}a + {p(1-q)(2-q)\over (p+2)~2!}a^2 +...]
\ee

The second regularization is 
\be
\int _0^1~dx (x^2+a^2)^{p-1\over 2} ((1-x)^2+a^2)^{q-1\over 2} \equiv B(p,q;a)
\ee

We would like to expand in powers of $a$ maintaining the duality symmetry between $p$ and $q$.
To this end we generalize the usual relation between Beta and Gamma functions. Consider
		 the ``regularized'' Gamma function.
				 \be
\int _0^\infty ~dx ~e^{-x}(x^2+ a^2)^{p-1\over 2} \equiv \tilde \Gamma (p,a)
				 \ee

Using the change of variables $x=x'y'$ and $x+y=y'$, we can write 
\[
\tilde \Gamma (p,a) \tilde \Gamma (q,a) ~=~
\int _0^\infty ~dx ~e^{-x}(x^2+ a^2)^{p-1\over 2} 
\int _0^\infty ~dy ~e^{-y}(y^2+ a^2)^{q-1\over 2} 
\]
\[
= \int _0^\infty dy'~ y' e^{-y'} 
\int _0^1 dx'~ (y'^2x'^2 + a^2 )^{p-1\over 2} (y'^2 (1-x')^2 + a^2 ) ^{q-1\over 2}
\]
\[
= \int _0^\infty dy'~ (y')^{p+q-1} e^{-y'} 
\int _0^1 dx'~ (x'^2 + ({a\over y'})^2 )^{p-1\over 2} ((1-x')^2 + ({a\over y'})^2 ) ^{q-1\over 2}
\]
\be     \label{GGB}
= \int _0^\infty dy'~ (y')^{p+q-1} e^{-y'} 
B(p,q;{a\over y'})
\ee

Thus using (\ref{GGB}), if we have an expansion for $\tilde \Gamma (p;a)$ in powers of $a$ we can deduce the expansion of 
$B(p,q;a)$.

We write 
\[
\int _0^\infty ~dx ~e^{-x}(x^2+ a^2)^{p-1\over 2} \equiv \tilde \Gamma (p,a) = \tilde \Gamma _1(p,a)+ \tilde \Gamma _2 (p,a) 
\]
\[
\tilde \Gamma _1 (p,a) = \int _0^{la} ~dx ~e^{-x}(x^2+ a^2)^{p-1\over 2}
\]
\be
\tilde \Gamma _2(p,a) = \int _{la}^\infty ~dx ~e^{-x}(x^2+ a^2)^{p-1\over 2} 
\ee
We have introduced an arbitrary parameter $l$ ($l>1$). The exponential in $\tilde \Gamma _1$ can be expanded
in powers of $x$ because the series converges uniformly in the interval $(0,la)$. In 
$\tilde \Gamma _2$, $(x^2 + a^2)^{p-1\over 2} $ can be expanded in powers of $1/x$ because again the series
converges uniformly for $l>1$ . The dependence on the arbitrary parameter $l$ should cancel order by order 
in the sum $\tilde \Gamma_1 +\tilde \Gamma _2$. We will verify this to the order that we calculate.

\be
\tilde \Gamma _1 = \sum _{n=0}^\infty \int _0 ^{la} {(-x)^n \over n!} [x^2 + a^2]^{p-1\over 2}
\ee 
Consider
\be
\int _0^{la} x^n (x^2 + a^2)^{p-1\over 2} dx 
\ee

The change of variable $x=ax'$ followed by $t=x'^2+1$ gives
\be
{a^{n+p}\over 2}\int _1^{l^2+1} {dt \over \sqrt {t-1}}(t-1)^{n\over 2} (t)^{p-1\over 2}
\ee
A further change $s={1\over t}$ brings it into a standard form:
\[
{a^{n+p}\over 2}\int _{1\over l^2+1} ^1 ds (1-s)^{{n+1\over 2}-1}s^{{n+p\over 2}-1}
\]
\be
={a^{n+p}\over 2} [
B(-({n+p\over 2}),{n+1\over 2})
-B_{1\over l^2+1}(-({n+p\over 2}),{n+1\over 2})]
\ee

The leading $n=0,1$ piece gives
\[
\tilde \Gamma _1(p,a) = {a^p\over 2} [B( -{p\over 2},{1\over 2}) - B_{1\over {l^2+1}} (-{p\over 2}, {1\over 2})]  - {a^{1+p}\over 2}[B(-({1+p\over 2}),1)]+...
\]
\be
=-{a^p\over p} + {a^{p+1}\over p+1} +
 {(al)^p\over p}  + {(al)^p\over 2l^2}{p-1\over p-2} +  {(al)^p\over l^4} {(p-1)(p-3)\over (p-4)} +...
\ee

Here we have kept the $l$-dependent terms from the $n=0$ piece and only the $l$-independent piece from $n=1$.

\[
\tilde \Gamma _2 = 
\int _{la}^{\infty} dx e^{-x}x^{p-1}[1+ {a^2\over x^2}]^{p-1\over 2}  
\]
\[
=\int _{la}^{\infty} dx e^{-x}[x^{p-1} + a^2{p-1\over 2} x^{p-3} + {a^4\over 2!}({p-1\over 2})({p-3\over 2}) x^{p-5} +...
\]
\[
= \Gamma (p,la) + a^2{p-1\over 2}\Gamma (p-2,la) + {a^4\over 2!}({p-1\over 2})({p-3\over 2})\Gamma (p-4,la) +...
\]
Here $\Gamma (\alpha , x)$ is the incomplete Gamma function \cite{GR}. It is defined as
\[
\Gamma (\alpha ,x) = \int _x^\infty ~dt e^{-t}t^{\alpha -1}
\]
It has the power series expansion:
\[
\Gamma (\alpha ,x) = \Gamma (\alpha ) - \sum _{n=0}^{\infty} {(-1)^n x^{\alpha +n} \over n! (\alpha +n)}
\]
Thus $\tilde \Gamma _2$ becomes
 
\be
=\Gamma (p) - {(al)^p\over p}  - {(al)^p\over 2l^2}{p-1\over p-2} -  {(al)^p\over l^4} {(p-1)(p-3)\over (p-4)} +...
\ee
Adding we see that the $l$-dependent terms cancel to the order they have been calculated, and the result is :
\be
\tilde \Gamma (p,a) = \Gamma (p) - {a^p\over p} + {a^{p+1}\over p+1} +...
\ee

Plugging all this into (\ref{GGB}) we get for the ``regularized'' Beta function: 
\be
B(p,q;a) = B(p,q) - {a^p\over p} - {a^q\over q} + a^{p+1}{(q-1)\over (p+1)} + a^{q+1} {(p-1)\over (q+1)}+...
\ee
Interestingly enough this agrees with the first regularization scheme to this order.

\end{document}